# Financial Portfolios based on Tsallis Relative Entropy as the Risk Measure

Sandhya Devi[1]
Edmonds, WA, 98020, USA
Email: sdevi@entropicdynamics.com

**Abstract:** Earlier studies have shown that stock market distributions can be well described by distributions derived from Tsallis entropy, which is a generalization of Shannon entropy to non-extensive systems. In this paper, Tsallis relative entropy (TRE), which is the generalization of Kullback-Leibler relative entropy (KLRE) to non-extensive systems, is investigated as a possible risk measure in constructing risk optimal portfolios whose returns beat market returns. Portfolios are constructed by binning the risk values and allocating the stocks to bins according to their risk values. The average return in excess of market returns for each bin is calculated to get the risk-return patterns of the portfolios. The results are compared with those from three other risk measures: 1) the commonly used 'beta' of the Capital Asset Pricing Model (CAPM), 2) Kullback-Leibler relative entropy, and 3) the relative standard deviation. Tests carried out for both long (~18 years) and shorter terms (~9 years), which include the dot-com bubble and the 2008 crash periods, show that a linear fit can be obtained for the risk-excess return profiles of all four risk measures. However, in all cases, the profiles from Tsallis relative entropy show a more consistent behavior in terms of both goodness of fit and the variation of returns with risk, than the other three risk measures.

## 1. Introduction

In capital asset management, risk optimal portfolios are usually based on using the covariance $\beta$ defined in modern portfolio theory [1] and the capital asset pricing model (CAPM) [2][3][4][5][6] or simply the standard deviation $\sigma$ as volatility measures. These measures are based on the efficient market hypothesis [7][8] according to which a) investors have all the information available to them and they independently make rational decisions using this information, b) the market reacts to all the information available reaching equilibrium quickly, and c) in this equilibrium state the market has a normal distribution. Under these conditions, the return for an equity $j$ is linearly related to the market return [9] as

[1] Shell International Exploration and Production Co. (Retired)

$$R_j = \beta_j\, R_m + \alpha_j \tag{1a}$$

$\beta_j$ is the risk parameter given by

$$\beta_j = \rho_{j,m}\, (\sigma_j/\sigma_m) \tag{1b}$$

where $\rho_{j,m}$ is the correlation coefficient of $R_j$ and $R_m$, and $\sigma_j$ and $\sigma_m$ are the standard deviations of $R_j$ and $R_m$.

The intercept $\alpha_j$ is the value of $R_j$ when $R_m$ is zero and hence can be considered as the excess return of the equity above the market return. The return $R$ over a period $\tau$ is defined as

$$R(t,\tau) = (X(t) - X(t-\tau))/X(t-\tau) \tag{1c}$$

$X(t)$ is the stock value at time $t$.

In 1972, empirical tests of the validity of CAPM were carried out by Black, Jensen and Scholes [10] who examined the monthly returns of all the stocks listed in NYSE for 35 years, between 1931-1965. Portfolios were constructed by binning the estimated risk parameter $\beta$ and allocating the stocks for each bin according to their risk parameter. The long term (35 years) results showed a highly linear relationship between the excess portfolio return α and the bin risk parameter $\beta$, the slope being slightly positive. This indicates that the higher risk stock portfolios yield marginally higher excess returns. However, when the tests were carried out for shorter periods (~9 years), the relationship between the excess returns and $\beta$ were still linear but the slopes were non-stationary becoming even negative for some periods.

In reality, how true are the assumptions of CAPM? Observations show that the market is a complex system that is the result of decisions by interacting agents (e.g., herding behavior), traders who speculate and/or act impulsively on little news, etc. Such a collective/chaotic behavior can lead to wild swings in the system, driving it away from equilibrium into the regions of nonlinearity. Further, the stock market returns show a more complicated distribution than a normal distribution. They have sharper peaks and fat tails (Figure 1).

Hence, there is a need to define a risk measure which is not bound by the constraints of CAPM. There have been several publications which argue that entropy is one such risk measure. In statistical mechanics, entropy is a measure of the number of unknown microscopic configurations of a thermodynamical system that is consistent with the measurable macroscopic quantities such as temperature, pressure, volume, etc. It is a measure of the uncertainty in the system [11][12]. In 1948, Shannon applied the concept of entropy as a measure of uncertainty to information theory, deriving Shannon entropy [13]. In finance, there are several features which make entropy more attractive as a risk measure. It is more general than the standard deviation [14][15] since it depends on the probabilities. Depending on the type of entropy used, it is capable of capturing the non-linearity in the dynamics of stock returns [16]. A review of applications of entropy in finance can be found in [17].

There have been several empirical studies comparing the predictive power of Rényi and Shannon entropies [18][19] with those from other measures (in particular $\beta$ and σ) with respect to portfolio expected returns. The conclusions are [19] that in the long run, the risk optimal portfolios from both Rényi and Shannon entropies show significantly lower variance than those from either σ or $\beta$.

In this work, we investigate the use of Tsallis relative entropy (TRE) [20], Kullback-Leibler relative entropy (KLRE) [21], beta, and relative standard deviation as risk measures for constructing risk optimal portfolios and compare the results with those of CAPM. In the CAPM tests by Black, Jensen and Scholes [10], the portfolio returns and risk parameters are defined relative to the market returns and risks. Hence, any new risk measures to be tested and compared with CAPM results also must be relative.

Several studies [22][23] indicate that the issues connected with the assumptions of CAPM (viz. efficient market hypothesis) can be addressed using statistical methods based on Tsallis entropy [24], which is a generalization of Shannon entropy to non-extensive systems. These methods were originally proposed to study classical and quantum chaos, physical systems far from equilibrium such as turbulent systems (non-linear), and long range interacting Hamiltonian systems. However, in the last several years, there has been considerable interest in applying these methods to analyze financial market dynamics as well. Such applications fall into the category of econophysics [25].

The rest of the paper is organized as follows. In Section 2, Tsallis relative entropy with some necessary background on Tsallis entropy and $q$-Gaussian distribution is discussed. A relationship between TRE and the parameters of a $q$-Gaussian distribution is derived. Section 3 deals with the data and methodology for constructing risk optimal portfolios and their results. The conclusions are given in Section 4.

In this paper we use the terms volatility and risk interchangeably. Strictly speaking, the term volatility should be used since we only use the stock price time series for the analysis. However, in the literature the term risk has also been used to mean volatility.

The returns are calculated as defined in (1c). The term expected returns is used to mean predicted average returns.

## 2. Theory

### 2.1 Review of Tsallis Statistics

Tsallis entropy is a generalization of Shannon entropy

$$S_{sh} = \sum_i P_i \, ln(1/P_i) \tag{2}$$

to non-extensive systems. It is given by

$$S_q = \sum_i P_i \, ln_q(1/P_i) \tag{3}$$

where $P_i$ is the probability density function at the i[th] sample under the condition $\sum_i P_i = 1$ and the $q$ logarithm $ln_q(x)$ is given by

$$ln_q(x) = (x^{1-q} - 1)/(1-q) \tag{4}$$

The scaling parameter $q$ is a universal parameter, but its value can change from system to system.

Substituting (4) in (3), we get:

$$S_q = \left(1 - \sum_i P_i^q\right)/(q-1) \tag{5}$$

Unlike Shannon entropy, Tsallis entropy obeys a pseudo additive property

$$S_q(A+B) = S_q(A) + S_q(B) + (1-q)\, S_q(A)\, S_q(B) \tag{6}$$

The scaling parameter $q$ denotes the extent of the non-extensivity of the system. As $q \to 1$, the additive property of Shannon entropy is recovered.

Considering the continuous case for a random variable Ω, one can show [24] that the maximization of $S_q$ with respect to $P$ under the following constraints:

$$\int_{-\infty}^{\infty} P(\Omega) d\Omega = 1 \tag{7a}$$

$$\langle(\Omega - \overline{\Omega})\rangle_q = \int_{-\infty}^{\infty} (\Omega - \overline{\Omega})\, P^q(\Omega) d\Omega = 0 \tag{7b}$$

$$\langle(\Omega - \overline{\Omega})^2\rangle_q = \int_{-\infty}^{\infty} (\Omega - \overline{\Omega})^2\, P^q(\Omega) d\Omega = \sigma_q^2 \tag{7c}$$

gives the Tsallis $q$-Gaussian distribution:

$$P_q(\Omega) = \frac{1}{\hat{Z}}[1 + (q-1)b_1(\Omega - \overline{\Omega}) + (q-1)b_2(\Omega - \overline{\Omega})^2]^{1/(1-q)} \tag{8}$$

$\hat{Z}$ is the normalization. $b_1$ and $b_2$ are the Lagrange multipliers for the constraints (7b) and (7c) respectively. The expectation value $\langle - \rangle_q$ in (7b) and (7c) are the $q$–expectation values.

Equation (8) can be re-written in a $q$-Gaussian form

$$P_q(\Omega) = \frac{1}{Z_q}[1 + (q-1)B(\Omega - M)^2]^{1/(1-q)} \tag{9a}$$

$$M = \left(\overline{\Omega} - \frac{\mathrm{b}_1}{2b_2}\right) \tag{9b}$$

$$B = b_2 \Big/ \left(1 - (q-1)\frac{b_1^2}{4b_2}\right) \tag{9c}$$

The normalization $Z_q$ is given by

$$Z_q = \int [1 + (q-1)B(\Omega - M)^2\,]^{1/(1-q)}\, d\Omega \tag{10a}$$

$$= C_q \,/\sqrt{B}$$

$$C_q = \sqrt{\pi}\,\frac{\Gamma\left(\frac{1}{q-1} - \frac{1}{2}\right)}{\sqrt{q-1}\;\Gamma\left(\frac{1}{q-1}\right)} \tag{10b}$$

Here Γ is the gamma function. Note that in the limit $q \to 1$, it can be shown that the Tsallis entropy and the corresponding $q$-Gaussian distribution go to the Shannon entropy and the Gaussian distribution respectively.

As described in detail in an earlier publication [26], the parameters $q$, $B$ and $M$ are estimated using the method of Maximum Likelihood Estimation (MLE) [27]. For completeness, the MLE equations are given here. Denoting, for brevity

$$\varphi = 1/(q-1) \text{ and } \kappa = (q-1)B$$

The MLE equations for $\varphi$, $M$, and $\kappa$ are given by

$$[\psi(\varphi) - \psi(\varphi - ½)] = \frac{1}{N}\sum_i ln(1 + \kappa\Omega_i^2) \tag{11a}$$

$$M = \sum_i (w_i\,\Omega_i) \tag{11b}$$

$$½\,\kappa = \varphi\,\left(\frac{1}{N}\right)\sum_i(\,\Omega_i^2/(1 + \kappa\Omega_i^2\,) \tag{11c}$$

Here, $w = (1 + \kappa\,\Omega^2\,)^{-1}/\sum_i(1 + \kappa\,\Omega_i^2\,)^{-1}$ and $N$ is the number of samples. The weights $w$ are normalized.

Equations (11a) – (11c) are non-linear and have to be solved numerically. The details are given in [26]. We will denote the estimated values of $q$, $M$ and $B$ as $\hat{q}$, $\widehat{M}$ and $\hat{B}$ respectively.

It is easy to verify that for $q \to 1$

$$w \to 1$$

$$M \to \mu$$

$$B \to 1/(2\sigma^2)$$

$\mu$ and $\sigma$ being the mean and standard deviation respectively of $\Omega$.

### 2.2 Tsallis Relative Entropy

The generalization by Tsallis [20] of Kullback-Leibler relative entropy

$$S_{KL}(P \| R) = -\sum_i P_i \, ln \, (R_i/P_i) \tag{12}$$

to non-extensive systems is given by

$$S_T(P \| R) = -\sum_i P_i \, ln_q(R_i/P_i) \tag{13}$$

P and R are normalized PDF's.

Using the definition of $ln_q(x)$ given in (4),

$$S_T(P \| R) = (\sum P_i (P_i/R_i)^{q-1} - 1)/(q-1) \tag{14}$$

The following are some of the properties of $S_T$:

1. Asymmetry: $S_T(P \| R) \neq S_T(R \| P)$ (15)

2. Non-negativity [28]: Since $-ln_q(x)$ is a convex function for $q > 0$

$$S_T(P \| R) = -\sum_i P_i \, ln_q\left(R_i/P_i\right) \geq -ln_q\left(\sum_i P_i \left(R_i/P_i\right)\right) = 0 \tag{16}$$

3. Pseudo-additivity [28]:

$$\begin{aligned} S_T(P_1 + P_2 \| R_1 + R_2) = {} & S_T(P_1 \| R_1) + S_T(P_2 \| R_2) \\ & + (q-1) \, S_T(P_1 \| R_1) \, S_T(P_2 \| R_2) \end{aligned} \tag{17}$$

The first two properties hold for KL relative entropy as well.

Equation (17) shows the applicability of TRE to correlated systems. As $q \rightarrow 1$, the pseudo additivity becomes the additive property

$$S_{KL}(P_1 + P_2 \| R_1 + R_2) = S_{KL}(P_1 \| R_1) + S_{KL}(P_2 \| R_2)$$

## 2.3 Tsallis $q$-Gaussian Relative Entropy

Direct calculations of relative entropies defined in equations (12) and (13) in terms of histograms of the data have several problems:

1. They depend on the number of bins in the histograms.
2. The relative entropies are defined only in the overlapping region of $R$ and $P$.
3. $S_{KL}$ is finite only if $R$ is non-zero in all the overlapping bins. The same is true for $S_q$ but for $q$>1.

This makes the number of samples for the computation of relative entropies rather sparse and hence the stability and accuracy become questionable. However, if both $R$ and $P$ can be well fit with $q$-Gaussian distributions, analytical expressions for the relative entropies can be derived in terms of the parameters of these distributions. In an earlier work [26] we have shown that the financial market return distributions (S&P 500 and Nasdaq) can be well modelled by $q$-Gaussian distributions, even during the dot-com bubble and 2008 crash periods. If the distributions of the returns of individual equities can also be modelled by $q$-Gaussians, then we can derive analytical formulas for TRE of an individual equity $P$ with respect to the market $R$. Figure 2 shows monthly percentage returns of S&P 500 and five randomly chosen individual stocks (both from S&P 500 and Nasdaq) and the corresponding fit to $q$-Gaussian distributions. Visual inspection shows that the fits are pretty good. However, to quantify the 'goodness of fit,' Kolmogorov-Smirnov (KS) [29] tests are carried out. Briefly, this involves determining the maximum absolute distances $\mathbf{D_{max}}$ between the empirical and the synthetic $q$-Gaussian cumulative distribution functions (CDF). The fit is good if $\mathbf{D_{max}}$ is less than a critical distance $\mathbf{D_{crit}}$. The details of constructing synthetic $q$-Gaussian and determining $\mathbf{D_{max}}$ and $\mathbf{D_{crit}}$ are given in [26]. The values of $\mathbf{D_{max}}$ and $\mathbf{D_{crit}}$ for S&P 500 and the five randomly chosen stocks are displayed in Figure 2. In all cases $\mathbf{D_{max}}$ is < $\mathbf{D_{crit}}$ showing that the distributions of the returns of even the individual stocks can be modelled well with $q$-Gaussian distributions.

From equations (9) and (10), the Tsallis $q$-Gaussian distributions for the returns of a market index $R$ and an individual equity $P$ can be written as

$$R_q(\Omega) = \frac{1}{Z_{qR}}\left[1 + (q-1)B_R(\Omega - M_R)^2\right]^{1/(1-q)} \tag{18}$$

and

$$P_q(\Omega) = \frac{1}{Z_{qP}}\left[1 + (q-1)B_P(\Omega - M_P)^2\right]^{1/(1-q)} \tag{19}$$

The normalizations $Z_{qR}$ and $Z_{qP}$ are given by

$$Z_{qR} = C_q/\sqrt{B_R}$$

$$Z_{qP} = C_q/\sqrt{B_P}$$

The parameter $q$ is estimated from the reference distribution $R$. All three parameters $q$, $B_R$ and $M_R$ are estimated for $R$. For $P$, only $B_P$ and $M_P$ are estimated using $q$ of the reference distribution.

As shown in the Appendix, the Tsallis relative entropy is now given by

$$\begin{aligned} S_T(P \| R) &= -ln_q(\gamma_{RP}) \\ &\quad + \frac{1}{2}\gamma_{RP}{}^{1-q}[(\gamma_{RP}{}^2 - 1) + (3-q)B_R(M_P - M_R)^2] \end{aligned} \tag{20}$$

Here $\gamma_{RP} = \sqrt{B_R/B_P}$ .

In the limit $q \to 1$, $M \to \mu$, $B \to 1/(2\sigma^2)$, and $\gamma_{RP} \to \sigma_{RP}$, where $\sigma_{RP} = \sigma_P/\sigma_R$ is the relative standard deviation, giving the KL relative entropy

$$S_{KL}(P \| R) = -ln(\sigma_{RP}) + \frac{1}{2}(\sigma_{RP}^2 - 1) + (\mu_P - \mu_R)^2/(2\sigma_R^2) \tag{21}$$

$S_T(P \| R)$ is evaluated at the estimated parameters $\hat{q}$, $\hat{M}$ and $\hat{B}$. The first two moments are used to compute the $S_{KL}$.

Note that $S_T$ depends non-linearly on the returns since both $B$ and $M$ have non-linear dependence on the returns (equations (11b) and (11c)). The first two terms in (20) depend only on the generalized standard deviations. The third term is a distance term (complementary to correlation) in generalized average returns. This is a systematic risk as the one addressed in CAPM. Hence Tsallis relative entropy combines aspects of both the standard deviation and CAPM risk measures and addresses the non-linearity of the stock dynamics.

## 3. Data, Methodology, and Results

### 3.1 Data

For the present study, we consider daily stock data from 4 January 2000 to 30 May 2018. The reference market index is chosen to be the S&P 500. For portfolio construction, we consider securities in the S&P 500 as of 2018. The data are adjusted for dividends and splits. No attempt has been made to correct the data for inflation.

### 3.2 Methodology

In testing the performance of the four risk measures in this study (Tsallis relative entropy, Kullback-Leibler relative entropy, beta, and the relative standard deviation $\sigma_{RP}$), two procedures somewhat similar to that described by Black, Jensen and Scholes [10] are followed. The exact procedures are as follows.

Procedure I

We only consider securities in the S&P 500 as of 2018 which have data extending all the way back to January 1995. This gives us about 340 stocks to work with in Procedure I and this list of stocks remains the same for all the cycles in Procedure I.

Each cycle consists of the following three steps:

a) Five years of data prior to the starting date of the cycle (e.g., 4 January 2000 is the starting date of the first cycle) are used to estimate the parameters of the risk models for the reference market index S&P 500 and for each of the securities. The values of the relative entropy risk measures are calculated from the model parameters. The expected return is computed as the average monthly return (as defined in (1c)) of the security over the next six months (arbitrarily chosen).

b) The risk values are then binned and the securities are assigned to each bin according to their risk value such that there are an equal number of securities in each bin. Note that this makes the bin widths variable. The set of securities in each bin can be considered a portfolio. The risk value of each bin is taken to be the center value of the bin. The number of bins remains the same for all cycles.

c) Assuming an equal amount of money invested in every security, the expected return of the portfolio in each bin in excess of the S&P 500 expected return is calculated. This gives the risk-return values for each portfolio.

The data are then shifted by six months and steps a) - c) are repeated for the next cycle. The $q$ values, however, are estimated every year. The process is continued until all the data are exhausted. For the data period considered, this gives us 36 samples of average monthly returns for each security.

Note that every time the data are shifted, the contents of each bin in step b) can change (even though the number of stocks in each bin remains the same in Procedure I). Also, in step c), each bin is rebalanced every six months such that an equal amount of money is invested in every security. This means that if this procedure is applied in practice, some securities would be sold and others bought every six months to implement steps b) and c). The effects on the portfolio returns due to transaction costs incurred in such selling and buying and taxes imposed on realized gains are not included in this study.

Finally, the returns in each bin are further averaged over all the 36 samples of average monthly returns and the bin-risk values are also averaged over all cycles. This gives us the final risk-expected return profile. We denote the expected earnings of the portfolio in excess of expected market return as $E_{rel}$. Note that the binning procedure is expected to minimize the effect of estimation errors on the performance of the portfolios.

Procedure II

For each cycle, we use the securities in the S&P 500 as of 2018 which have data extending five years before the start date of the cycle, e.g., all the way back to January 1995 for the first cycle. As the cycles move forward in time, more and more securities enter the computation. This gives us about 340 stocks in the first cycle increasing to about 460 in the last cycle.

The rest of Procedure II is similar to Procedure I, except that the bin widths (determined in the first cycle) do not change with each cycle, but the number of securities in each bin can change with each cycle. This procedure is closer to that described by Black, Jensen and Scholes [10] than Procedure I where the number of securities in each bin are kept fixed.

### 3.3 Goodness of Fit

The performance of the four risk measures is assessed by estimating the $\chi^2$, which is one of the commonly used estimates in statistics [19] in determining the goodness of fit. This quantity shows how close the risk-return patterns are to a linear regression. If $\{s\}$ is a set of risk values of the bins and $\{e\}$ the corresponding portfolio earnings, then

$$\chi^2 = 1 - \frac{\sum_i [e_i - (p_0 + p_1 s_i)]^2}{\sum_i (e_i - \bar{e})^2} \tag{22}$$

Here $p_0$ and $p_1$ (intercept and slope) are the parameters of the linear fit and $\bar{e}$ is the mean of $e$. Note that the closer the values of $e$ to the linear fit, the closer is $\chi^2$ to 1.

### 3.4 Results

We will first discuss the results from Procedure I. Figure 3 shows the long term behavior of the $E_{rel}$ calculated from monthly returns vs. the risk for the four risk measures considered. The period is 2000-2018. Note that for this long period, the slopes of the linear fit in all four cases are positive, indicating that for greater relative risk there is greater relative return. This behavior is similar to that observed in the tests of the CAPM model [10] as well. The $\chi^2$ is comparable in all cases, with TRE giving a value somewhat better than those for the others and a higher slope.

Figure 4 shows the effect of increasing the number of securities in each portfolio (diversification) on the long term performance of the portfolios. In the case of TRE, we observe a consistent behavior of improved performance with an increase in the number of securities in the portfolio. However, this is not the case with the other risk measures. Further, in all cases, the goodness of fit of TRE is better than that of beta of the CAPM.

Tests of CAPM by Black, Jensen and Scholes [10] for shorter periods (~9 years) show that the linear relationship between risk and return is intrinsic and not the result of better statistics. However, the risk-return patterns are non-stationary, i.e., the slopes and intercepts vary widely for each period, the slopes becoming even negative in some cases. In the present work, we carry out similar tests of the four risk measures, dividing the data into two periods of 9 years each: a) 2000-2009 and b) 2009-2018. Note that the estimation of parameters starts from the data five years earlier: 1995-2004 for a) and 2004-2013 for b). Hence the first interval covers only the dot-com bubble period and the beginning edge of the 2008 crash. As shown in Figure 5, the $q$ values for period a) varies between 1.24 – 1.45 denoting a relatively calm situation. For interval b) the range is between 1.4 – 1.74 pointing to strong non-extensivity and possibly a chaotic situation. So unlike [10], these tests look at the intrinsic behavior of risk-return patterns during very different market conditions.

The risk-relative return profiles for the two 9 year periods are displayed in Figure 6. First of all, whereas TRE shows good $\chi^2$ values during both periods, the other three measures show rather poor $\chi^2$ values during the first period. Further, the profiles for $\beta$ show the biggest change in slope, going from negative in the first period to positive in the second. The other three risk profiles all show positive slopes, with that for TRE showing the least change in slope between the two periods. So these tests indicate that the TRE based profiles show a very consistent behavior even during periods that might include chaotic market situations.

We will now discuss the results from Procedure II. As discussed previously, in this case the number of stocks that enter the computation increases as time progresses. This means more number of stocks in each bin and hence better statistics and possibly better performance. Figure 7 shows the number of securities used for portfolio construction in each five-year window of step a) as a function of year.

The long term behavior of the $E_{rel}$ calculated from monthly returns vs. the four risk measures is shown in Figure 8. As in the previous procedure, the slopes of the linear fit in all four cases are positive, indicating that for greater relative risk there is greater relative return. Except in the case of $\beta$, all measures show a better $\chi^2$. However, of all the risk measures, TRE gives the best $\chi^2$ and highest slope. The profile with $\beta$ as the risk measure shows the worst $\chi^2$ and the smallest slope.

The results for the two shorter terms, 2000-2009 and 2009-2018, are shown in Figure 9. As in procedure I, the risk-return patterns of $\beta$ show the biggest change in slope going from negative in the first period to positive in the second. Such a swing also accounts for the poor performance in terms of both goodness of fit and returns (small slope) over the longer (18 year) period. The performance of TRE, however, is consistent for both the periods in terms of better $\chi^2$ values and higher positive slopes (higher returns for higher risk) compared to those from the other risk measures. This again indicates that the TRE based portfolios behave consistently during periods of different stock market characteristics.

## 4. Summary and Conclusions

In this work, we have proposed Tsallis relative entropy (TRE) as a novel risk measure for the selection of risk optimal portfolios for returns in excess of market returns. Since the distributions of the returns of both the market (S&P 500) and the individual stocks can be well fit with $q$-Gaussian distributions, the TRE can be analytically expressed in terms of the model parameters of the $q$-Gaussian distributions. This alleviates several problems (described in 2.3) encountered in the histogram based estimation of relative entropies. Further, the analytical expressions show that TRE has aspects of both the CAPM and the standard deviation risk measures in a non-linear way.

The performance of TRE as a risk measure is compared with those of three other risk measures: KLRE, beta of CAPM, and relative standard deviations. The KLRE is obtained as the limiting $q \rightarrow 1$ case of TRE.

One of the observations in these empirical tests is the consistent behavior of TRE. Over the long term, even though all risk measures show a linear relationship with earnings, TRE gives the best value of $\chi^2$ and highest slope. Over the shorter term consisting of periods of very different market characteristics (bubble and crash), the measures KLRE, $\beta$, and relative standard deviation show quite different behaviors. This is particularly obvious in the results of procedure II which has better statistics. The biggest change is shown by $\beta$ with slopes going from negative in the first period to positive in the second. Such a change in behavior degrades the long term performance as well in terms poor goodness of fit and returns. TRE, however, shows a consistent behavior in terms of good $\chi^2$ values and highest positive slopes, even during these shorter intervals. These results indicate that it might be possible to construct portfolios whose returns can beat the market return even during periods that might include chaotic situations such as bubbles and crashes, using TRE as the risk measure.

The empirical investigations in this work point to the importance of taking into account the non-linearity and correlations of stock market dynamics in defining risk measures. TRE is one such measure and may help in the construction of portfolios whose returns show a predictive risk-return behavior both in the long and shorter time investments.

This brings us to the question of how short is a ‘shorter time’ to hold the portfolio whose returns beat the market. That depends on the relaxation time of the market dynamics. In the case of Tsallis statistics, one might be able to get a handle on these by estimating $q_{triplet}$[24][30][31]. This, however, is a subject for future studies.

## Acknowledgements

Many thanks to Sherman Page for a critical reading of the manuscript.

## Appendix: Derivation of Tsallis q-Gaussian Relative Entropy

Denoting

$$\varphi = 1/(q-1) \text{ and } \kappa = (q-1)B \tag{A1}$$

the integral representation of Tsallis relative entropy (14) is given by

$$S_T(P\|R) = \varphi\left(\int P(P/R)^{1/\varphi} d\Omega - 1\right) \tag{A2}$$

with the PDF's $R$ and $P$ given by

$$R_q(\Omega) = \frac{1}{Z_{qR}}[1 + \kappa_R(\Omega - M_R)^2]^{-\varphi} \tag{A3}$$

$$P_q(\Omega) = \frac{1}{Z_{qP}}[1 + \kappa_P(\Omega - M_P)^2]^{-\varphi} \tag{A4}$$

Here, the normalizations

$$Z_{qR} = C_\varphi/\sqrt{\kappa_R} \tag{A5}$$

$$Z_{qP} = C_\varphi/\sqrt{\kappa_P} \tag{A6}$$

$$C_\varphi = \sqrt{\pi}\frac{\Gamma\left(\varphi - \frac{1}{2}\right)}{\Gamma(\varphi)} \tag{A7}$$

Using (A2) – (A6):

$$\int P\,(P/R)^{1/\varphi}\, d\Omega$$

$$= \left(\sqrt{\kappa_P/\kappa_R}\right)^{1/\varphi}\left(1/Z_{qP}\right)\int \frac{[1+\kappa_R(\Omega-M_R)^2]}{[1+\kappa_P(\Omega-M_P)^2]^{1+\varphi}}\, d\Omega$$

$$= \left(\sqrt{\kappa_P/\kappa_R}\right)^{1/\varphi}\left(1/Z_{qP}\right)\int \frac{[1+\kappa_R(M_P-M_R)^2]}{[1+\kappa_P(\Omega-M_P)^2]^{1+\varphi}} + \frac{\kappa_R(\Omega-M_P)^2}{[1+\kappa_P(\Omega-M_P)^2]^{1+\varphi}}\, d\Omega$$

Using:

$$\int 1/[1+\kappa_P(\Omega-M_P)^2]^{1+\varphi}\, d\Omega = Z_{qP}\left(\frac{\varphi-1/2}{\varphi}\right)$$

and

$$\int \kappa_R(\Omega - M_P)^2 / [1 + \kappa_P(\Omega - M_P)^2]^{1+\varphi} = Z_{qP}(\frac{1}{2\varphi})(\kappa_R/\kappa_P)$$

one can write, after some algebra:

$$S_T(P \| R) = -ln_q(\gamma_{RP}) + \frac{1}{2}\gamma_{RP}{}^{1-q}[(\gamma_{RP}{}^2 - 1) + (3-q)B_R(M_P - M_R)^2] \quad \text{(A8)}$$

In writing (A8), we have used (A1) and $\gamma_{RP} = \sqrt{B_R/B_P}$.

# Figures

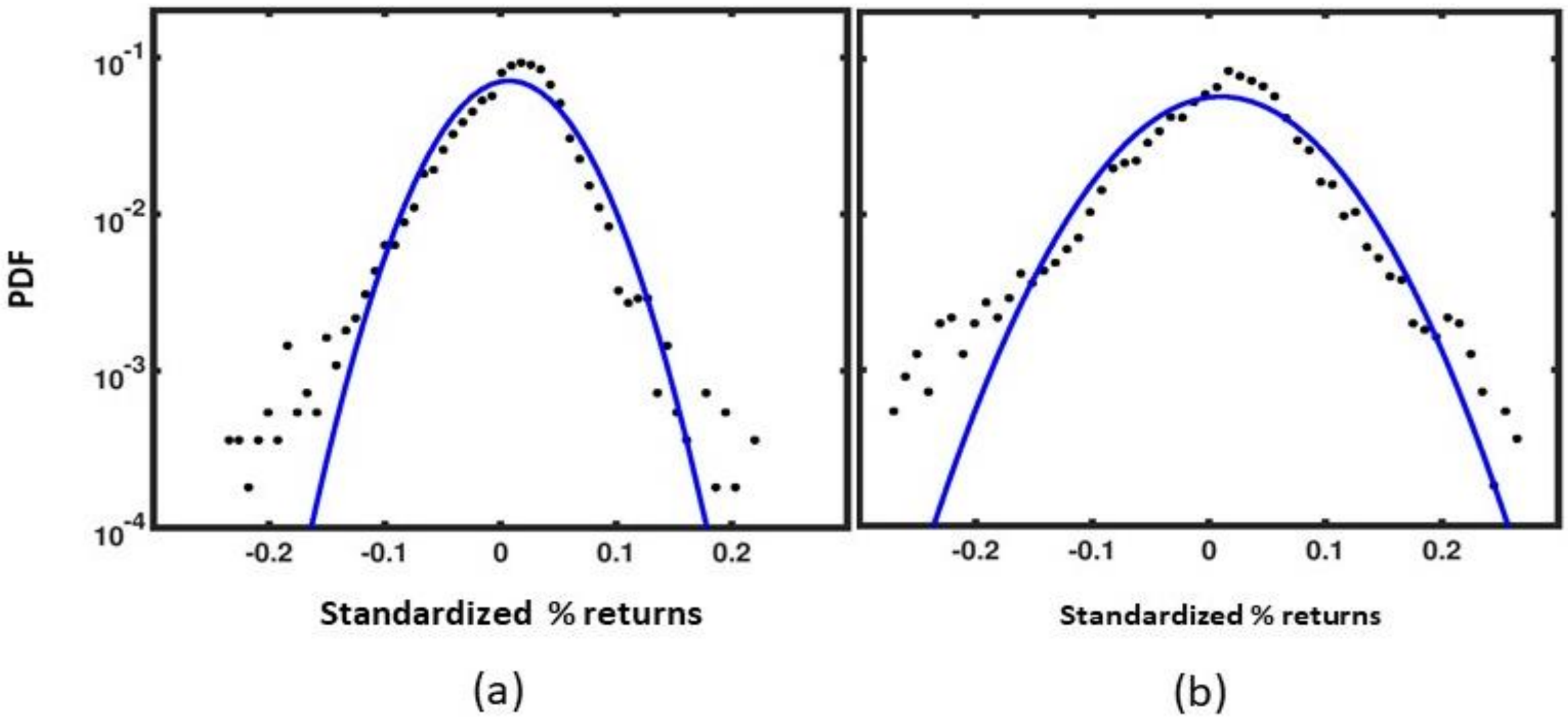

Figure 1. Comparison of the distributions of monthly standardized percent returns with the Gaussian distributions (solid blue line) having the same mean and standard deviation as the data (black dots). (a) S&P 500 for the period Jan 1995 – Jan 2017 and (b) Nasdaq over the same period.

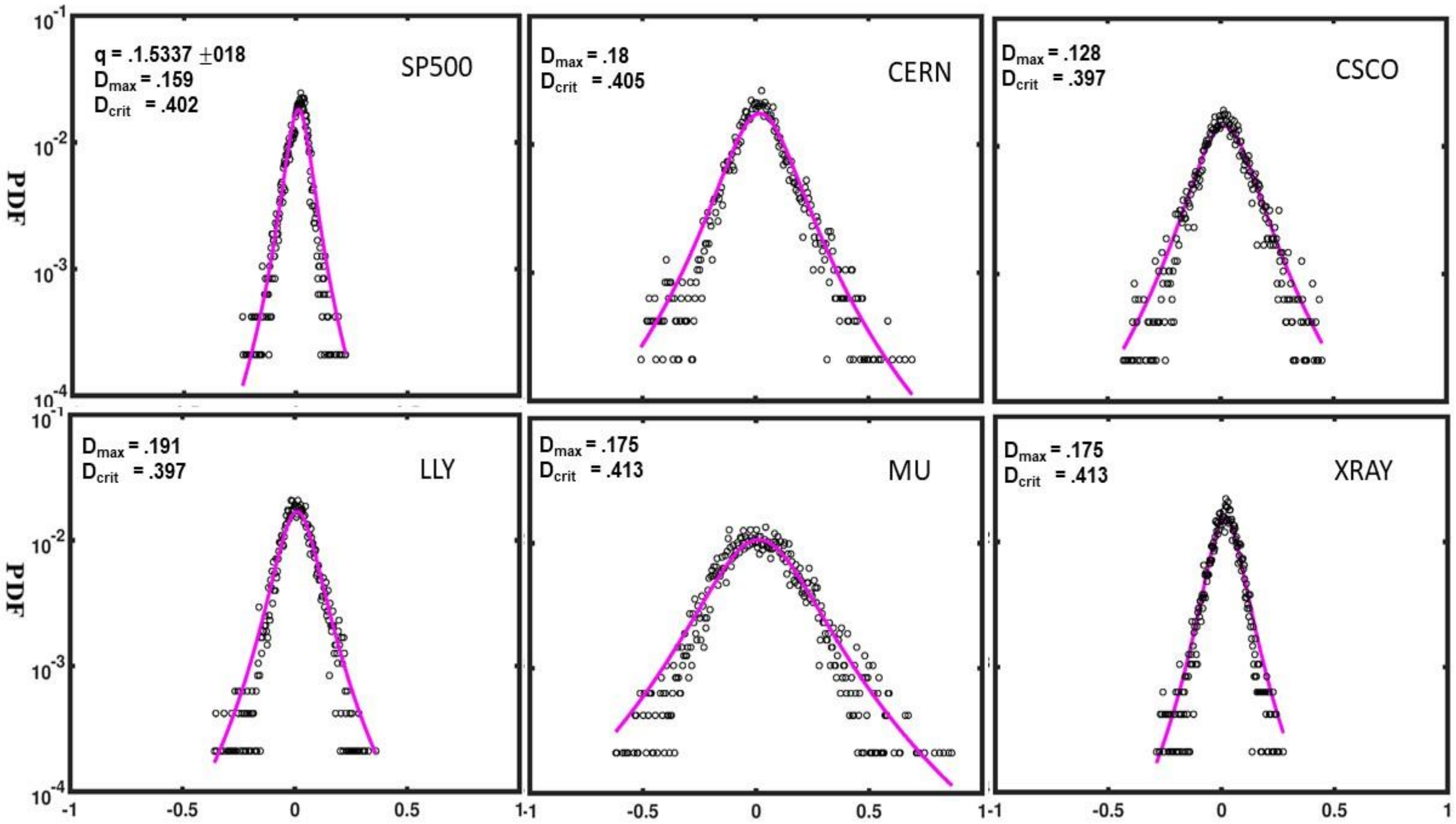

Figure 2. $q$-Gaussian fit to the distributions of monthly percent returns of S&P 500 and five randomly chosen stocks from S&P 500 and Nasdaq. The ticker symbols of the stocks are displayed on each corresponding figure. Period Jan 1995 – Jan 2017.

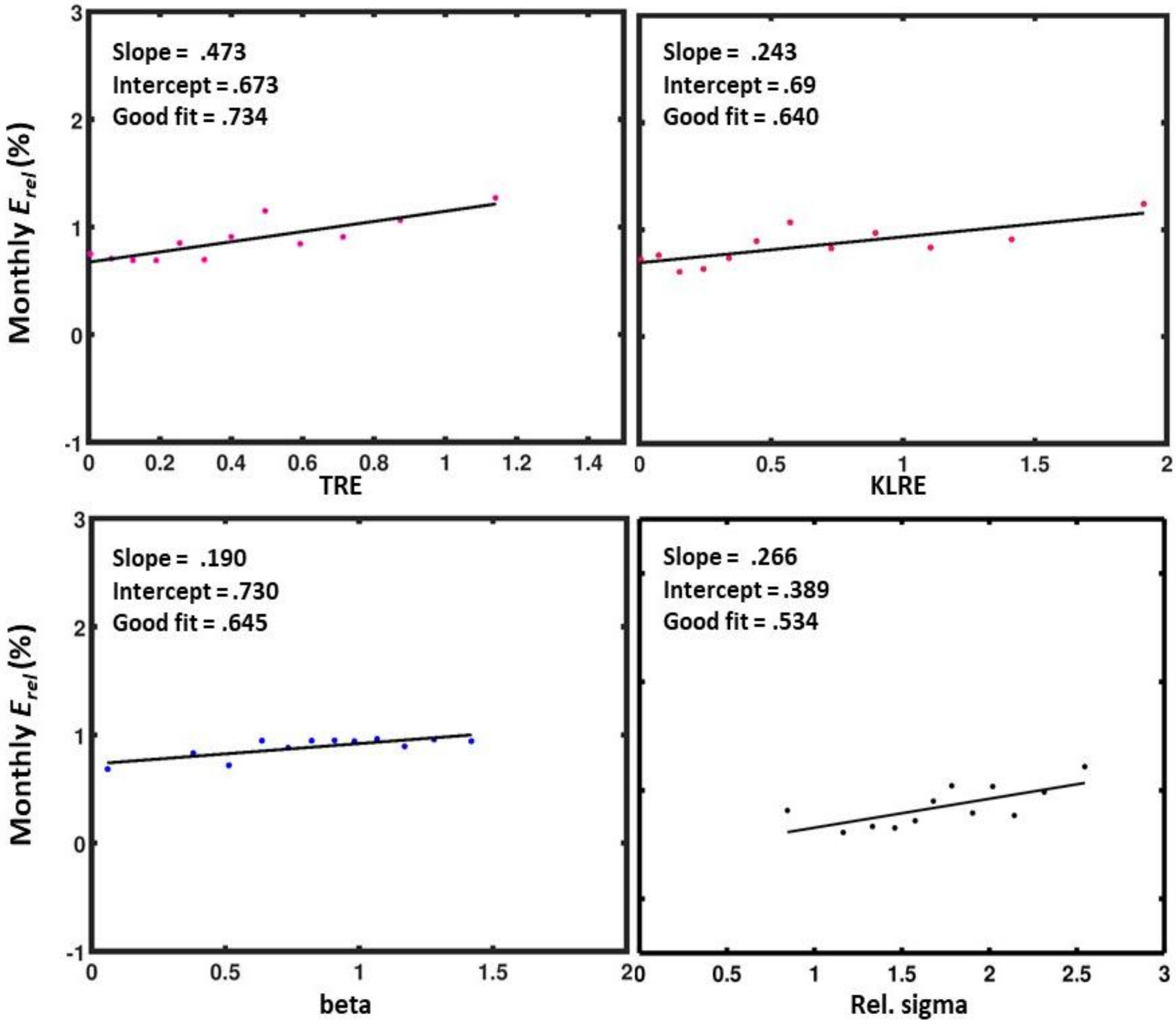

Figure 3. Average monthly excess returns of the portfolios vs. the four risk measures considered using Procedure I. The reference market index is the S&P 500. Portfolios constructed out of stocks in S&P 500. Number of stocks in each portfolio is 25. Data interval 2000-2018.

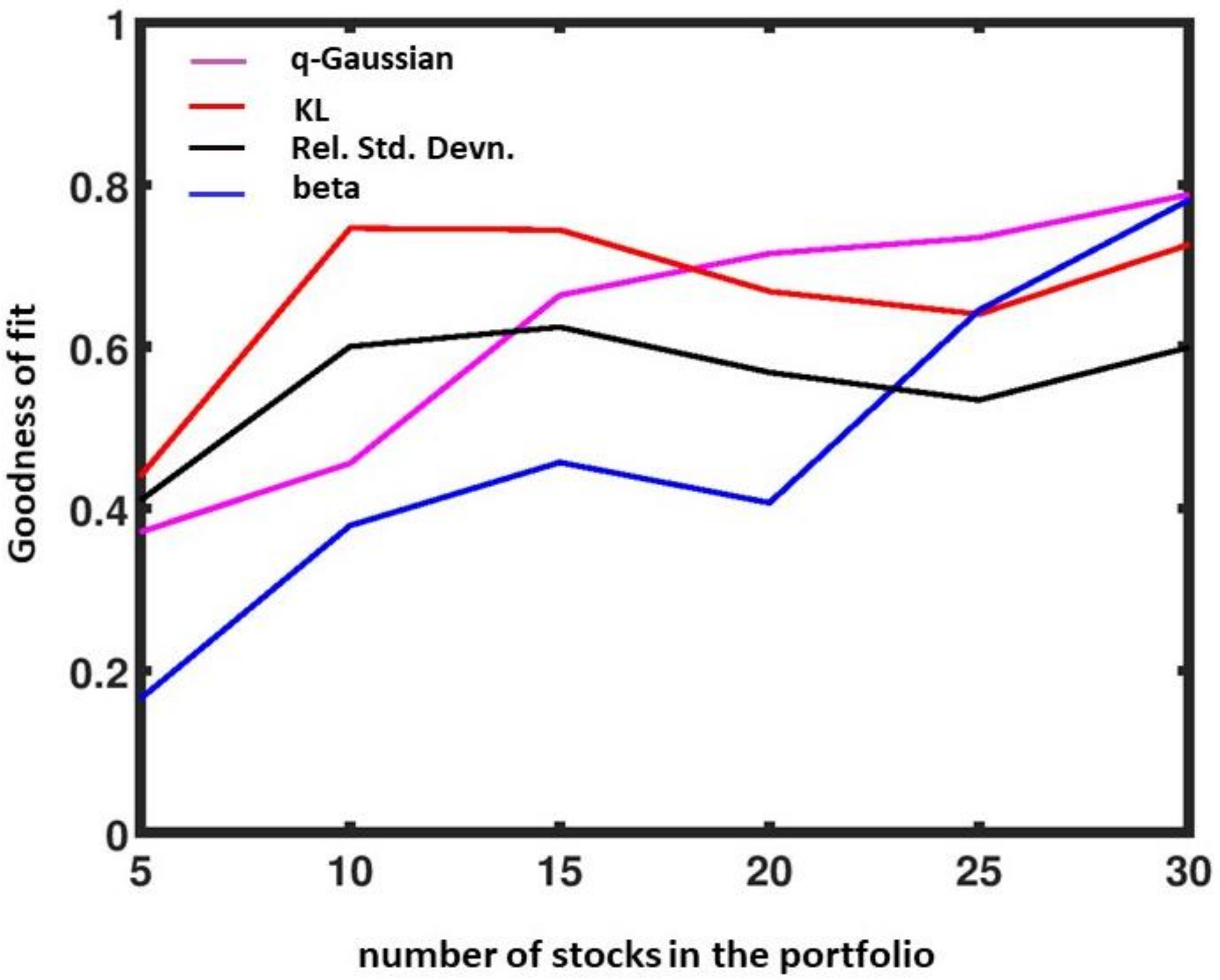

Figure 4. Performance of the four risk measures as a function of number of securities in the portfolios (diversification) for Procedure I.

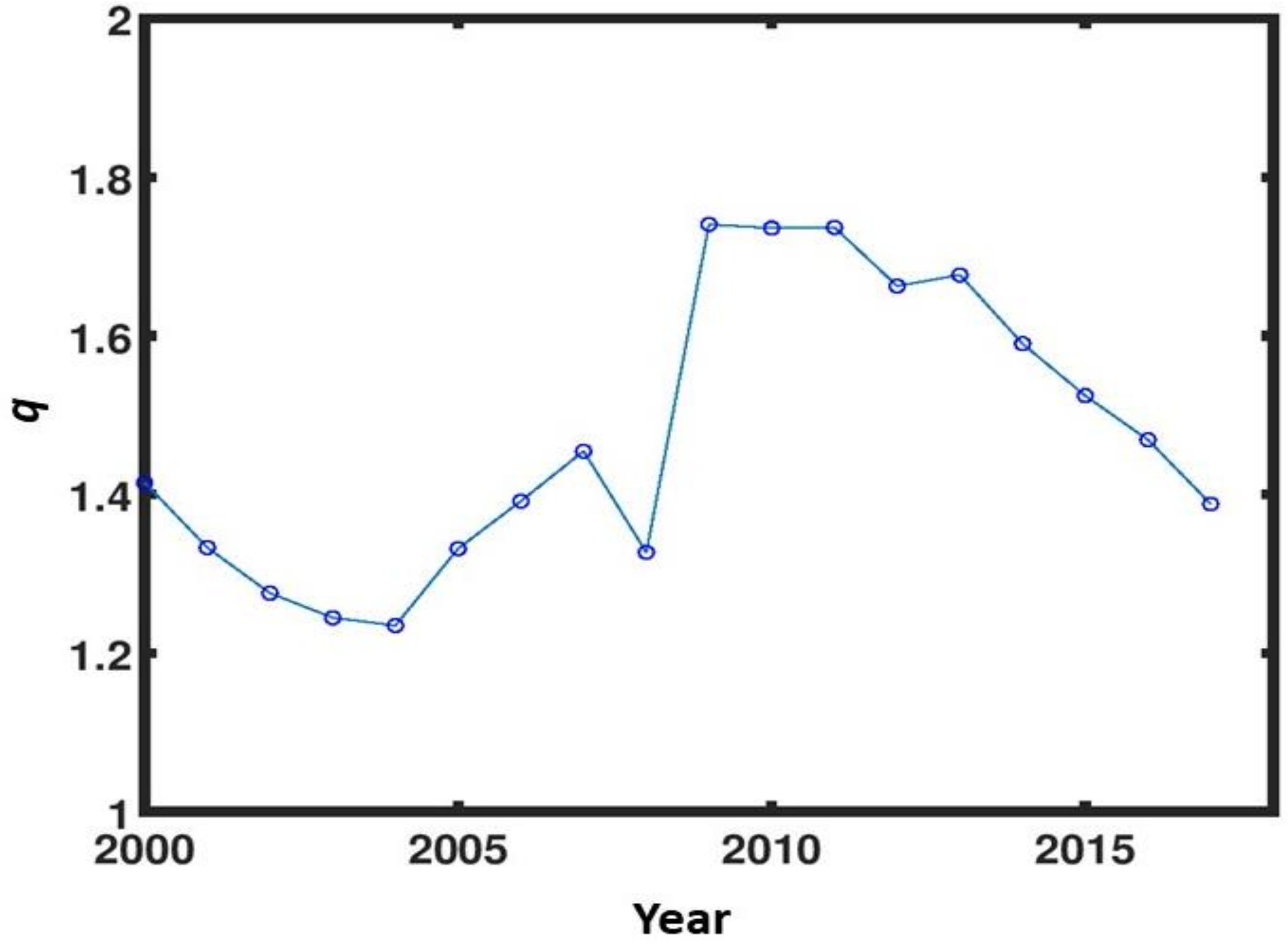

Figure 5. q variation over the investigation period 2000-2018 for Procedure I.

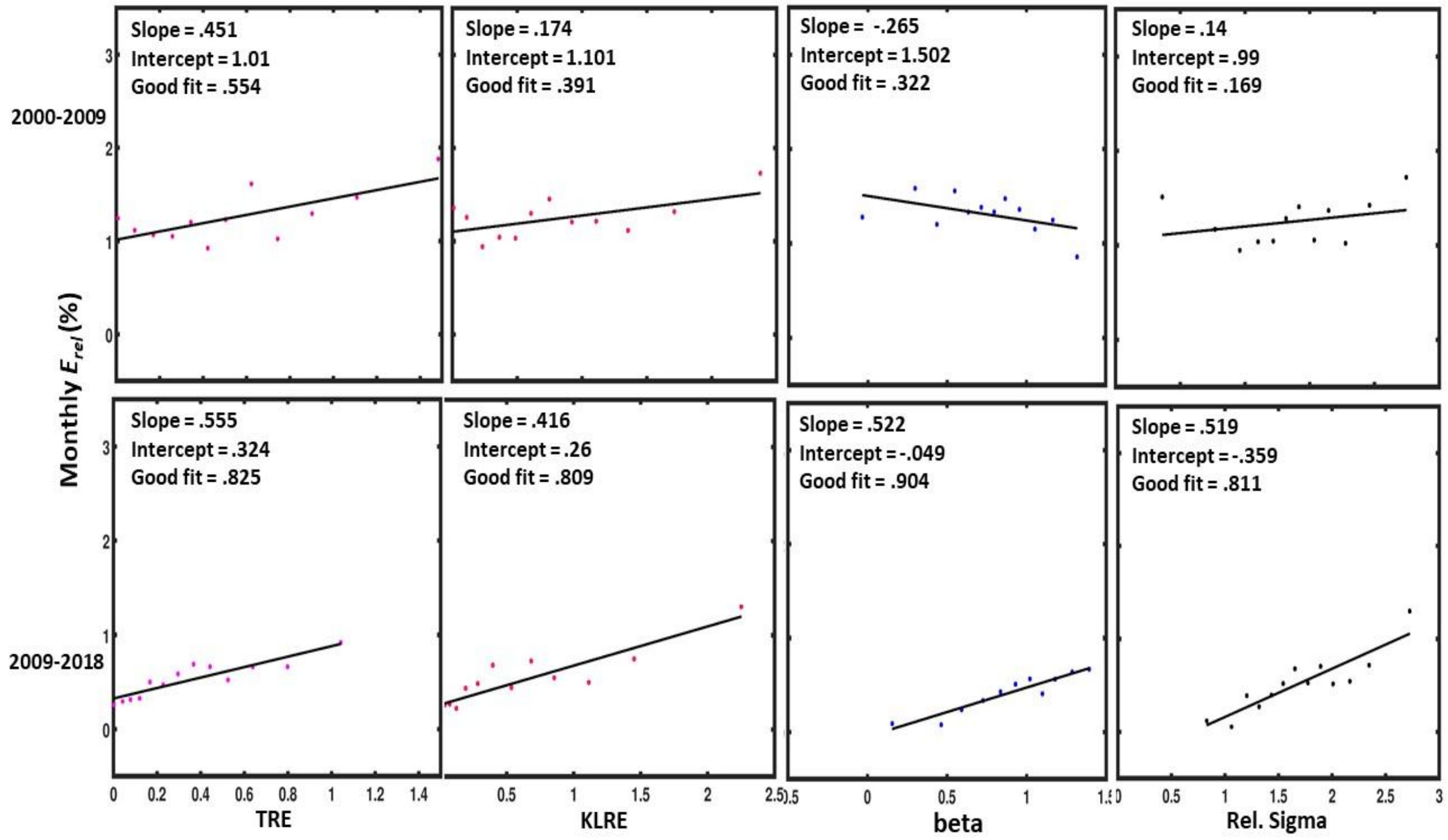

Figure 6. Comparison of average monthly excess returns of the portfolios vs. the four risk measures for shorter periods of 9 years, 2000-2009 and 2009-2018, for Procedure I.

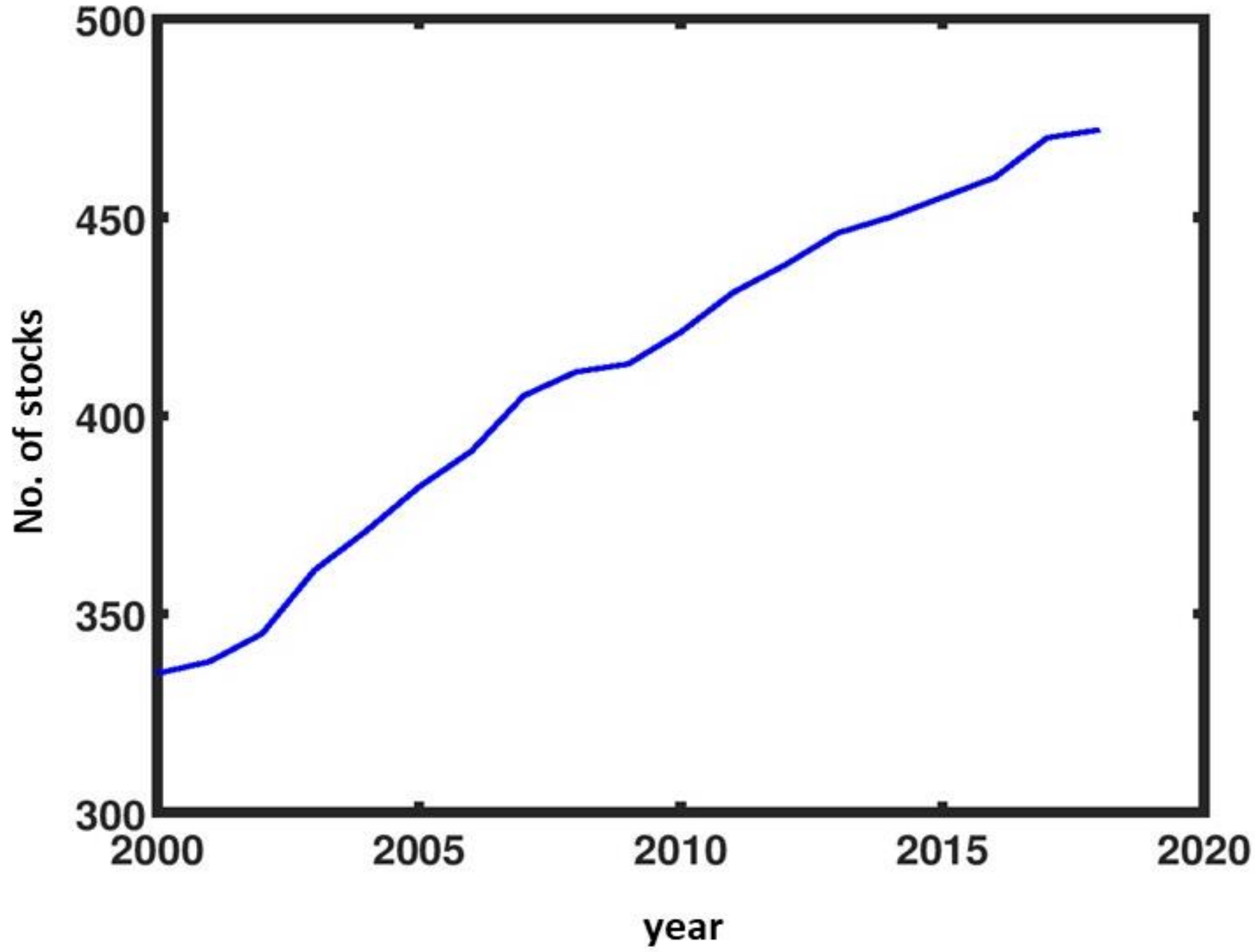

Figure 7. Number of stocks used for portfolio construction in each five-year window, plotted against the last year of each window, for Procedure II.

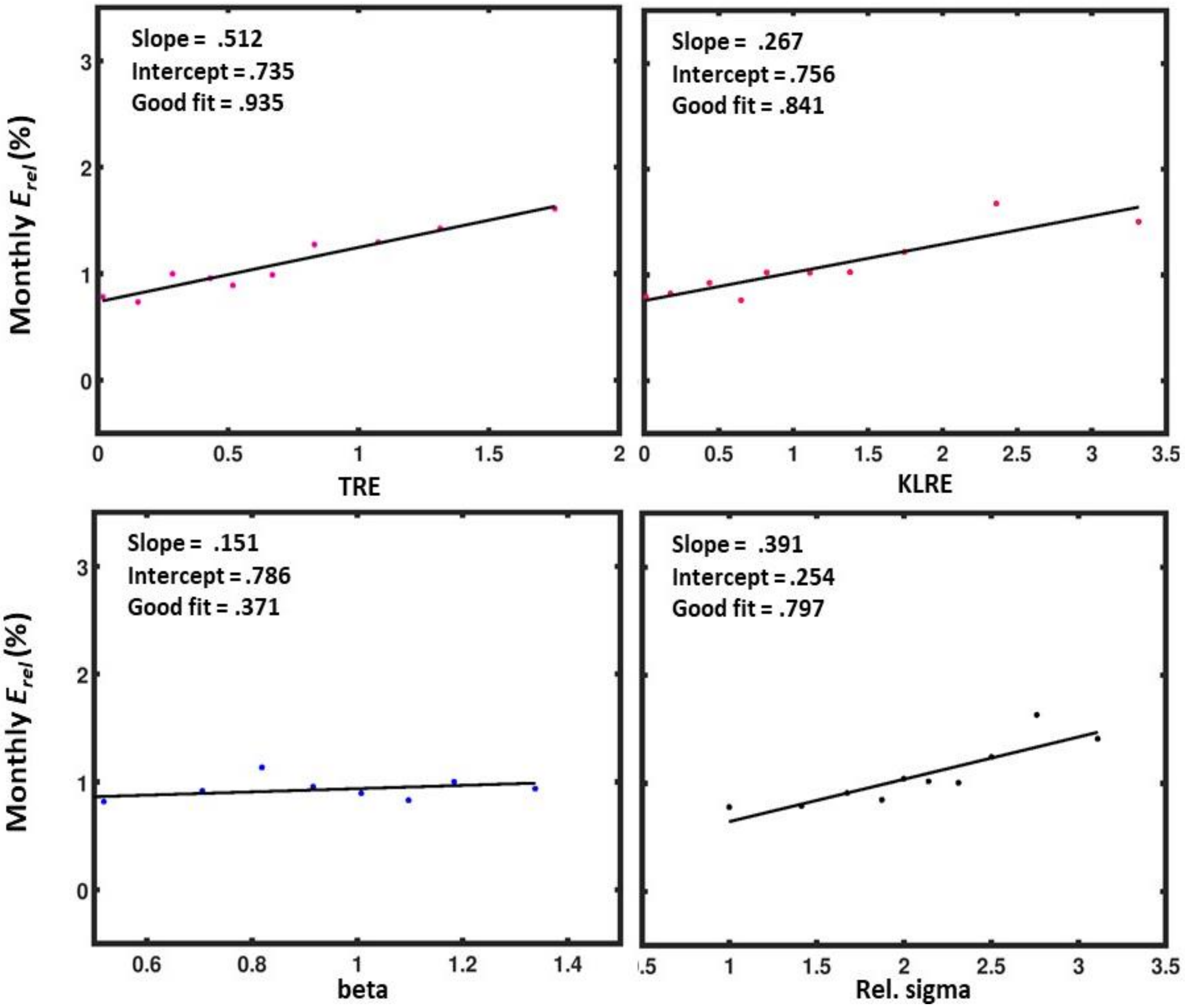

Figure 8. Same as Figure 3 for Procedure II.

.

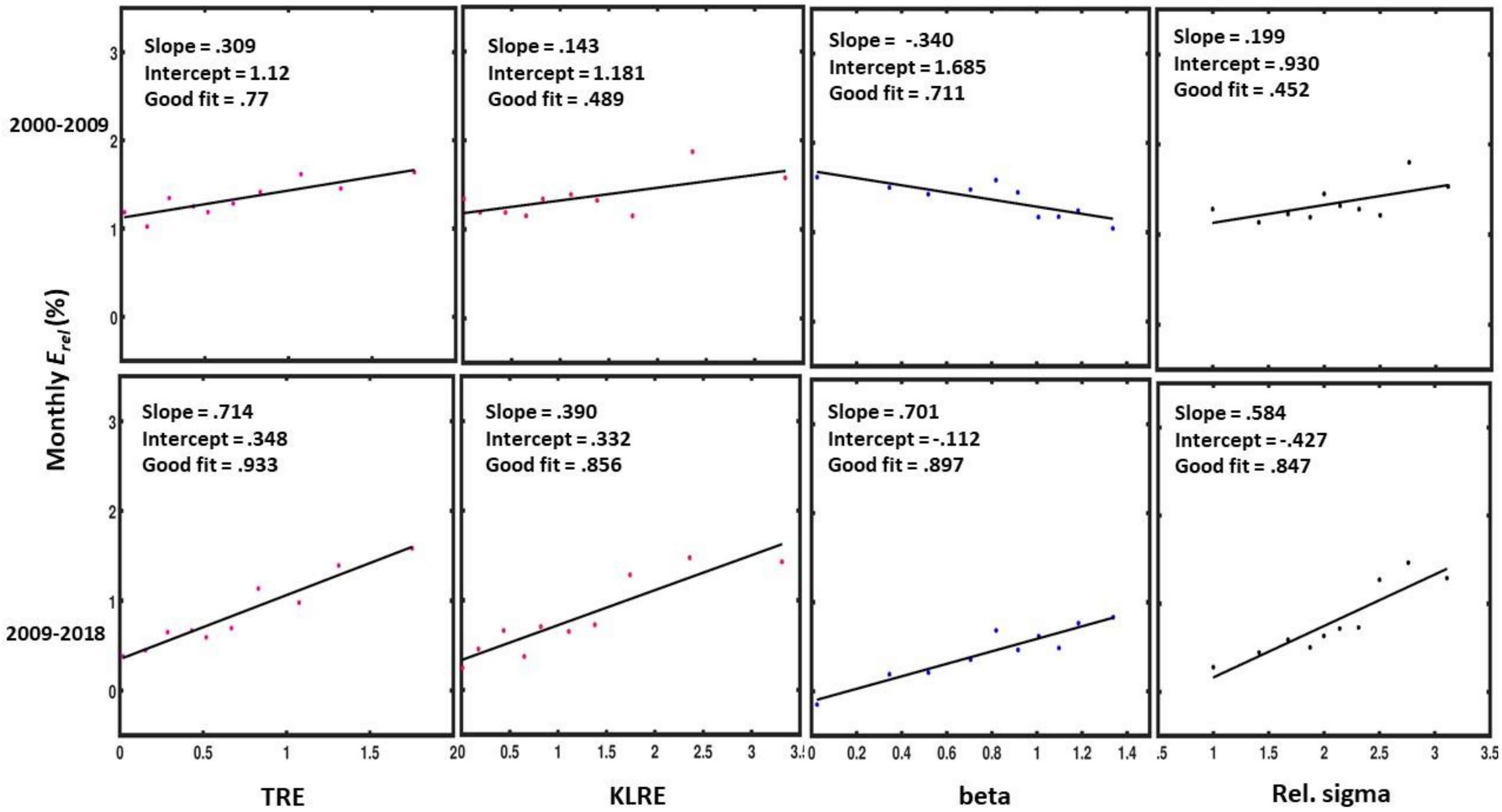

Figure 9. Same as Figure 6 for procedure II.